\documentclass[fleqn,usenatbib]{mnras}

\usepackage{newtxtext,newtxmath}

\usepackage[T1]{fontenc}

\DeclareRobustCommand{\VAN}[3]{#2}
\let\VANthebibliography\thebibliography
\def\thebibliography{\DeclareRobustCommand{\VAN}[3]{##3}\VANthebibliography}

\usepackage{graphicx}	
\usepackage{amsmath}	
\newcommand{\solmass}[1]{#1~M$_\odot$}

\newcommand{\rev}[1]{\textcolor{black}{#1}}

\title[Disc tearing in GW Ori]{High-resolution simulations of disc tearing in the GW Orionis triple system}

\author[A. K. Young]{
Alison K. Young$^{1,2,3}$\thanks{E-mail: a.k.young@leeds.ac.uk (AKY)}
\\
$^{1}$ School of Physics and Astronomy, University of Leeds, Sir William Henry Bragg Building, Woodhouse Ln, Leeds LS2 9JT, UK \\
$^{2}$ Centre for Exoplanets and Habitability, University of Warwick, Coventry CV4 7AL, UK \\
$^{3}$ Department of Physics, University of Warwick, Coventry CV4 7AL, UK 
}

\date{Accepted XXX. Received YYY; in original form ZZZ}

\pubyear{\the\year{}}

\begin{document}
\label{firstpage}
\pagerange{\pageref{firstpage}--\pageref{lastpage}}
\maketitle

\begin{abstract}
The disc around the pre-main-sequence triple star system GW Orionis is known from observations to be warped and broken. Theoretical modelling has produced conflicting results regarding the mechanism responsible for breaking the disc. Analytical predictions for the measured parameters of GW Ori suggest the disc is only marginally stable to tearing. We present new high-resolution simulations of GW Ori that replicate the wavelike regime expected in thick, low-turbulence protoplanetary discs for the first time to settle this question. Using the most optimistic values of misalignment and stellar mass ratio allowed by observational constraints, we find that the GW Ori disc can be torn by stellar torques alone, without need for an embedded planet. Even if the disc retains a smooth warp in simulations with similar parameters, it is likely that any small perturbation in the density or temperature structure could cause the disc to break. The new simulations rule out retrograde disc rotation relative to the stellar orbits and tentatively suggest the thicker ($h/r = 0.04$) disc better matches observations. Going forward, we should take care to ensure models of GW Ori and similar systems appropriately represent the propagation of warps. Additionally, analytical predictions are derived from idealized \rev{(and often massless)} discs and it is useful to assess how each observed disc might deviate from those assumptions\rev{, especially in the context of a young and active star-forming neighbourhood}.
\end{abstract}




\section{Introduction}
\label{sec:intro}
The triple pre-main-sequence stellar system GW Orionis has received much attention over the last decade, initially due to its unusual light curve and extended disc \citep[e.g.][]{fang2014,mathieu1995}, and more recently after observations revealed the inner disc to be shaped by disc tearing \citep{kraus2020}.

Disc tearing is a phenomenon that results from precession torques acting on a disc \citep{nixon2012}. A circumbinary disc or circumstellar disc around one component of a binary will develop a warped structure if it is not coplanar with the stellar orbits due to the radial dependence of the precession torques. If the warp becomes steep enough, the disc may become unstable to breaking. Physically, this occurs when the timescale for the communication of the warp becomes greater than the precession timescale \rev{and the stellar torques can no longer be communicated throughout the disc}. In protoplanetary discs, warps are expected to be communicated as bending waves so the wave travel time across the whole disc $t_w \approx 2 r_{\rm {out}} / c_s$ is the relevant timescale \citep{nixon2013}, \rev{where $c_s$ is the local sound speed. The disc thickness is determined by the sound speed (temperature) so is key in determining the disc evolution. The time scale for solid body precession of a circumbinary disc is given by
\begin{equation}
\omega_{\rm p} = \frac{3}{4} \frac{M_2}{M_1 + M_2} \frac{a^2}{r_{\rm{out}}^2} \omega_{\rm disc} \cos\Phi, \\
t_{\rm p} = \left | \frac{2\pi}{\omega_{\rm p}} \right |,
\label{eq:tprec}
\end{equation}
\citep{ivanov1999,bate2000oct,nixon2013}, where $\omega_{\rm disc}$ is the angular frequency of the disc at radius $r_{\rm{out}}$.
Hence, the evolution of a misaligned disc depends on several parameters including the misalignment between the disc and stellar orbital plane $\Phi$, the mass ratio of the binary components $M_1$ and $M_2$, and binary separation $a$. 
The binary eccentricity $e$ can also contribute strongly to the precession torque on the disc and can quickly drive large misalignments \citep{aly2015}. All these factors affect whether a disc can be torn by stellar torques \citep[see e.g.][]{fragner2010,nixon2013,facchini2013,aly2015,dogan2015,young2023,rabago2024}.} Large binary-disc misalignments of $\Phi \ge 40^\circ$ are generally required but an eccentric binary may tear a disc with a far smaller misalignment \rev{due to the additional precession torque}.

\rev{So far we have described the situation where the warp propagates as bending waves, which is known as the {\it wavelike} regime. This occurs where the Shakura-Sunyaev shear viscosity is smaller than the disc semi-thickness: $\alpha_{\rm{ss}} \ll h/r$ \citep{shakura1973,papaloizou1995,lubow2000}, i.e. typically for thick, low-viscosity discs like protoplanetary discs. When $\alpha_{\rm{ss}} \gg h/r$, the warp is communicated by internal viscous torques and the disc evolves in the {\it viscous} or {\it diffusive} regime. Highly ionised accretion discs around black holes are an example of the latter regime. Viscous discs tear when the precession torque is stronger than the viscous torque, which could occur close to accreting black holes \citep{nixon2012,nixon2013}.}

GW Orionis is a hierarchical triple star system that hosts an unusually massive and extended circumtriple disc \citep{mathieu1991,mathieu1995,berger2011,czekala2017}. High-resolution observations revealed a misaligned and eccentric inner disc, attributed to tearing by stellar torques \citep{kraus2020,bi2020}. \citet{kraus2020} presented smoothed particle hydrodynamical (SPH) simulations showing that the GW Ori disc is torn by the torques of the central triple star system. However, questions have since been raised as to whether this is the case or whether a giant planet first carved a gap in the disc, separating the inner and outer disc into independently precessing components. The embedded planet hypothesis was motivated by SPH simulations of GW Orionis that produced warped discs that did not break or only tore when the disc was probably unrealistically thin \citep{bi2020,smallwood2021,smallwood2025}. \citet{young2023} presented high--resolution simulations that indicated that a disc with similar misalignment to the GW Ori system could be susceptible to tearing without a planet-carved gap. The combination of a misalignment angle of $28^\circ$ \footnote{\rev{The value of $38.5^\circ$ given in \citet{kraus2020} was an unfortunate typographical error but the simulations presented there employed the correct rotations based directly on the observationally-derived angles.}} and the moderate eccentricity of the tertiary component $e_{\rm c} \approx 0.4$ likely puts the GW Ori disc on the borderline of stability. The previous simulations by \citet{kraus2020,bi2020,smallwood2021,smallwood2025} may have had insufficient resolution to accurately recover wavelike warp propagation because of excess numerical viscosity (see \citealt{drewes2021}). Consequently, some of the simulated discs were likely able to maintain smooth warp structures or instead tore in the viscous regime \footnote{Note that the simulations presented in \citet{smallwood2021} use a disc misalignment angle of $38^\circ$ which is the inclination of the GW Ori disc \emph{on the sky} \rev{ and outwith the observational uncertainties.}}.

Another consideration is the interpretation of the observationally-derived parameters. The longitude of the ascending node of the binary and disc position angle must be included in calculation of misalignment, in addition to the difference in inclination. When simulating GW Ori and other systems, care should be taken to note the observational uncertainties. In the case of GW Ori, there is sufficient uncertainty in the derived parameters to allow for an increased misalignment and reduced binary mass ratio that would increase the likelihood of disc tearing \citep{young2023}.

The aim of this work is to generate new high-resolution simulations of the GW Ori disc, verifiably in the wavelike regime, to determine whether the stellar torques can break the disc without the need, for example, of an embedded planet. In acknowledgement of the observational uncertainties, the most extreme values of system parameters are chosen within the allowed ranges. If disc tearing by the stellar torques alone can be ruled out under these conditions, then the inference that another mechanism must be at play is robust.

In this paper, the details of the simulation parameters and initial conditions are first detailed in section \ref{sec:method}. Next, the resolution of the simulations is investigated in section \ref{sec:validity} and we demonstrate that the simulated discs are wavelike. The results showing the discs tearing for prograde and retrograde discs are presented in sections \ref{sec:prograde} and \ref{sec:retrograde}. In section \ref{sec:discussion}, the simulations are compared with the observationally--derived properties of GW Ori and the reasons for discrepancies with prior modelling are discussed. 

\section{Method}
\label{sec:method}
Three high-resolution smoothed particle hydrodynamics (SPH) simulations were performed with {\sc phantom} \citep{price2018aa} to investigate the evolution of the GW Ori disc with $h/r = 0.03$ and $h/r = 0.04$ with prograde rotation, and with $h/r = 0.05$ and retrograde rotation (relative to the stellar orbits). The $h/r = 0.04$ simulation ran for more than 2 months on 48 cores before the disc broke. It was therefore unfeasible to simulate a prograde disc with $h/r = 0.05$ because it would take even longer to break - if it does indeed become unstable.

\subsection{Initial conditions}
\label{sec:initconds}

The stellar system is approximated as a binary, replacing the inner binary with a single star of the combined mass.  \citet{smallwood2021} demonstrated through n-body simulations that this is a reasonable approximation for GW Ori. The relevant parameters are therefore those of the outer binary (AB-C) and the \rev{circumtriple} disc. The orbital parameters are taken from the observationally-derived data of \citet{kraus2020}. We select values for the parameters at the optimistic end of their ranges allowed by the observational uncertainties to maximise the likelihood of the disc breaking. The parameters used to set up the stellar orbits and disc are listed in table~\ref{tab:orbitalparams}, including the observational uncertainties for the relevant parameters. The stars are modelled as sink particles with accretion radii set to 2.0 and 0.2~au for the sinks representing AB and C components respectively. These values allow accretion streams to form within the disc cavity but prevent the simulation being slowed significantly by SPH particles on very small orbits.

\begin{table}
 \caption{Initial parameters for setting up the circumbinary disc simulations. $a$,$\omega$, $i$, $\Omega$,and $e$ are the conventional orbital elements. Other parameters are described in the text. For the parameters chosen at the optimistic end of the range, the observationally-derived values and uncertainties are given. *Values for orbit AB-C in \citet{kraus2020}.}
    \label{tab:orbitalparams}
    \centering
    \begin{tabular}{ccc}
        Binary & Simulation & \citet{kraus2020} \\
        \hline 
     $M_{\rm AB}$     & \solmass{3.39} & \solmass{$2.47 \pm 0.33$}~$+$~\solmass{$1.43 \pm 0.18$} \\
     $M_{\rm C}$    &  \solmass{1.64} & \solmass{$1.36\pm 0.28$}\\
     $r_{\rm{acc},AB}$ & 2.0~au & \\
     $r_{\rm{acc}, C}$ & 0.2~au &  \\
     $a$      &  8.93~au & $8.89\pm 0.04$~au*\\
     $\omega$ &  105$^\circ$ & \rev{$105^\circ \pm 1^\circ$*} \\
     $i$      &  150.3$^\circ$ & $149.6^\circ \pm  0.7$* \\
     $\Omega$ &  232$^\circ$ & $230.9^\circ \pm 1.1$*\\
     $e$      &  0.382 & $0.379 \pm 0.003$*\\
     \hline \hline
     Disc & \\
     \hline
     $r_{\rm{in}}$ & 19~au &\\
     $r_{\rm{out}}$ & 150~au &\\
     $\Sigma(r_{\rm{in}})$ & $91.6$~g~cm$^{-2}$& \\
      $p$ & 1.5 & \\
      $q$ & 0.5 & \\
    \end{tabular}
   
\end{table}

The disc is modelled with $10^7$ SPH particles and is initialised with the location of the inner edge informed by the extent of the observed inner cavity in submillimetre \rev{continuum} observations\footnote{\rev{It would be preferable to derive the inner edge from gas emission or scattered light, which traces small grains that follow the gas distribution. However, the resolution of the CO emission map is far lower than the continuum map and in the scattered light observations the inner disc edge is hidden by the coronagraph.}} \citep{bi2020,kraus2020}. \rev{Given the detections of circumbinary material and substantial accretion onto the central stars \citep{calvet2004,fang2014,kraus2020}, the inner edge should be close enough to the stars for material to be disrupted and transferred inwards.} It is not computationally feasible to model the whole $\gtrsim 400$~au \citep{fang2017} disc so it is necessary to truncate the disc at 150~au. We discuss the potential limitations of this in section \ref{sec:uncertainty}. 
The disc surface density is set via 
\begin{equation}
\label{eq:surfdens}
    \Sigma (r) = \Sigma(r_{\rm {in}}) (r/r_{\rm in})^{-p},
\end{equation}
with $p=1.5$. The surface density profile is chosen to be steep to maximise the mass (and therefore also resolution) in the inner disc where the precession torques are strongest. \citet{young2023} showed that the amplitude of a warp wave decays as it propagates outwards through a disc with a shallow $p=0.5$ surface density profile, but with $p=1.5$ the warp amplitude is roughly conserved. This means a disc with a steeper surface density profile is more likely to break.
The mass of the GW Ori disc is measured to be \solmass{0.12} for $r<242$~au \citep{fang2017}. Integrating equation \ref{eq:surfdens} over $19 < r <242$~au gives the surface density normalisation $\Sigma(r_{\rm {in}})$ (Table~\ref{tab:orbitalparams}). 
The sound speed profile is given by 
\begin{equation}
    \label{eq:csprofile}
    c_{\rm s} (r) = c_{\rm s}(r_{\rm {in}}) (r/r_{\rm in})^{-q},
\end{equation}
with $q=0.5$. The initial sound speed normalisation $c_{\rm s}(r_{\rm {in}})$ is found from the disc thickness $h/r$.

The initial disc inclination relative to the $x-y$ plane is based on the derived inclination (on the sky) of Ring R1, $i_{\rm disc}=142\pm1^\circ$ \citep{kraus2020}. The face-on view of the simulations is then equivalent to the view of GW Ori on the sky. \rev{The longitude of the ascending node $\Omega_{\rm disc} = 180\pm8^\circ$ was derived from the same observations. The values of  $i$ and $\Omega$ are chosen to maximise $i_{\rm disc} - i_{\rm AB-C}$ and $\Omega_{\rm disc}- \Omega_{\rm AB-C}$ such that the misalignment,
\begin{equation}
\label{eq:misalignment}
    \Phi = \cos{i_{\rm disc}}\cos{i_{\rm AB-C}} + \sin{i_{\rm disc}}\sin{i_{\rm AB-C}}\cos(\Omega_{\rm disc} -\Omega_{\rm AB-C} ),
\end{equation}
 is the maximum value allowed by observations.}
The chosen values of inclination $i$ and longitude of the ascending node/position angle $\Omega$ shown in tables \ref{tab:orbitalparams} and \ref{tab:sims} give a misalignment of $\Phi = 34^\circ$ for the prograde discs and $\Phi = 110^\circ$  for the retrograde disc. 

\subsection{Simulation parameters}

The simulations were performed with {\sc phantom} and evolved with a locally isothermal equation of state. We employ the viscosity switch of \citet{cullen2010} to vary the artificial viscosity in the range $0.001 < \alpha_{\rm{SPH}} < 1.0$, and $\beta_{\rm{SPH}}=2.0$. The reason for this choice of artificial viscosity implementation is not to mimic a Shakura-Sunyaev viscosity explicitly \citep[e.g.][]{lodato2010}, but to achieve the lowest level of dissipation possible since protoplanetary discs are expected to be nearly inviscid with $\alpha_{\rm{ss}} \lesssim 10^{-3}$ \citep[e.g.][]{rosotti2023}. In the following section we determine the equivalent $\alpha_{\rm{ss}}$ in the simulations.

\begin{table}
  \caption{Summary of simulations conducted: inclination $i$ (not binary-disc misalignment), position angle $\Omega$, disc semi-thickness $h/r$, and the corresponding misalignment angle $\Phi$.}
    \centering
    \begin{tabular}{c|cccc}
        Name & $i$ [$^\circ$] & $\Omega$ [$^\circ$] & $h/r$ & $\Phi$ [$^\circ$]  \\
        \hline
         h/r = 0.03 & 141 & 172 & 0.03 & 34\\
         h/r = 0.04 & 141 & 172 & 0.04 & 34\\
         Retro & 52 & 172 & 0.05 & 110 \\
    \end{tabular}
  \label{tab:sims}
\end{table}

\section{Results}

In this section, we first verify that the simulations are in the wavelike regime. The evolution of the simulated discs is then presented, with further analysis given in \ref{sec:discussion}.

\subsection{Model validity}
\label{sec:validity}

\begin{figure}
    \centering
    \includegraphics[width=0.9\columnwidth]{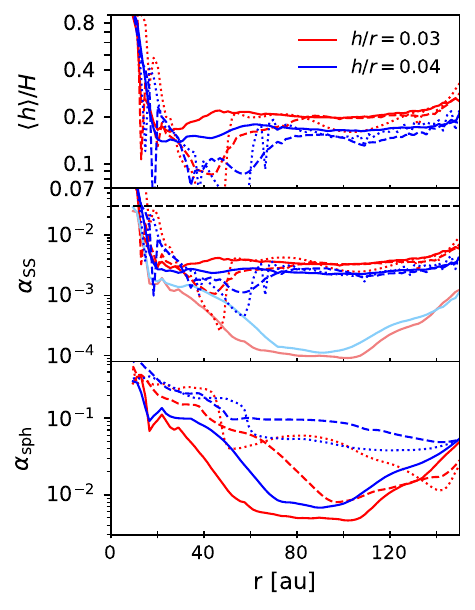}
    \caption{The resolution of the $h/r=0.03$ and $h/r=0.04$ simulations: ring-averaged values of $\left < h \right >/H$ (top), $\alpha_{\rm{ss}}$ (middle), and $\alpha_{\rm{SPH}}$ (bottom). Solid lines: snapshot as warp develops, dashed lines: snapshot as disc starts to break, dotted lines: after disc has torn- the same snapshots as shown in Fig.~\ref{fig:densplots}. In the middle panel, the pale red and pale blue lines give the $\alpha_{\rm{ss}}$ estimate for the same first snapshots but using the \citet{lodato2010} formulation (equation \ref{eq:lodatoalpha}). The dashed black line shows where $\alpha_{\rm{ss}}= 0.03 = h/r$.}
    \label{fig:resolution}
\end{figure}

Resolving the wavelike regime is crucial to obtaining accurate evolution \rev{so we require that $\alpha_{\rm{ss}} \ll h/r$ (see section \ref{sec:intro}).} Therefore we ensure that the equivalent viscosity generated by the SPH artificial viscosity is consistent with this regime. 

The shear viscosity can be estimated from the artificial viscosity parameters and the ring-averaged smoothing length $\left < h \right >$ following \citet{meru2012}:

\begin{equation}
\label{eq:alpha_ss}
     \alpha_{\rm ss} \approx \frac{31}{525}\alpha_{\rm SPH}\left(\frac{\langle h \rangle}{H}\right) + \frac{9}{70\pi}\beta_{\rm SPH}\left(\frac{\langle h \rangle}{H}\right)^2.
\end{equation}

The constants are calculated for the SPH kernel, which in this case is the M$_4$ cubic spline. Both the $\alpha_{\rm SPH}$ and $\beta_{\rm SPH}$ viscosity terms contribute to the shear viscosity. If the disc is well-resolved (i.e. $\left < h \right > \ll H$), the quadratic $\beta_{\rm SPH}$ term is small but it can become dominant in poorly resolved regions, especially since typically $\beta_{\rm SPH} = 2.0$.

To construct Fig.~\ref{fig:resolution}, \rev{smoothing lengths and $\alpha_{\rm SPH}$ of the particles} were averaged in radial bins. The scale height $H$ for each radial bin is estimated from the average particle height above the disc midplane\rev{, which is equivalent to $H$ since both assume a Gaussian vertical density profile}. The upper panel of Fig.~\ref{fig:resolution} shows the vertical resolution. The discs are resolved with at least $\sim5 $ smoothing lengths per scale height in most of the disc.
Fig.~\ref{fig:resolution} (lower panel) shows the estimated values of $\alpha_{\rm ss}$ for snapshots from the prograde disc simulations. Throughout most of the disc, $\alpha_{\rm ss} \lesssim 0.1 h/r$, given $h/r = 0.03$ or 0.04 initially\footnote{\rev{The scale height does evolve slightly but does not decrease below the initial value for at any radius.}}, and we are firmly in the wavelike regime including when the discs break (dashed lines) at the breaking radius ($\sim40$ and $\sim52$ au respectively).

For comparison with other works, we also estimate the shear viscosity using the \rev{commonly used} approach of \citet{lodato2010}:
\begin{equation}
    \label{eq:lodatoalpha}
    \alpha_{\rm ss} \approx 0.1 \alpha_{\rm SPH}\frac{\langle h \rangle}{H} .
\end{equation}
\rev{This was derived as part of a study that determined how SPH numerical viscosity can be used to represent a Shakura-Sunyaev disc viscosity, building on earlier studies of the physical nature of the SPH viscosity formulation \citep[including][]{artymowics1994,murray1996,lodatopringle2007}.} 

The pale red and blue lines in Fig.~\ref{fig:resolution} (middle panel) show $\alpha_{\rm{ss}}$ estimates using equation \ref{eq:lodatoalpha} for prograde disc simulation snapshots. Using equation \eqref{eq:lodatoalpha}, the estimated $\alpha_{\rm ss}$ viscosity is substantially underestimated, by an order or magnitude in much of the disc. \rev{This difference is mostly due to the absence of the quadratic term, which \citet{meru2012} later showed to be significant in many cases.}
\rev{Less significant contributions to the difference are due to the SPH formulation for which equation \ref{eq:lodatoalpha} was derived. \citet{lodato2010} used an early version of {\sc phantom} in which} artificial viscosity was applied to both approaching and receding particles whereas disc models in the \rev{current version of {\sc phantom} (and in {\sc sphNG)}} apply artificial viscosity to approaching particles only. Additionally, \rev{ the earlier version of {\sc phantom} implemented} a slightly different averaging of the artificial viscosity parameters\footnote{Compare equations 39 and 40 of \citet{price2018aa} with equations 24 and 25 of \citet{lodato2010}.}. Equation \ref{eq:alpha_ss} should be used with the standard version of {\sc phantom}.

A viscosity switch is employed to minimise the numerical viscosity and the bottom panel of Fig.~\ref{fig:resolution} shows the ring-averaged values of $\alpha_{\rm SPH}$. The switch increases $\alpha_{\rm SPH}$ for approaching particles to reproduce shocks accurately and we can see that this becomes relevant for the inner disc regions ($\lesssim 60 $ au).

The analysis here demonstrates that the simulations satisfactorily reproduce the wavelike regime at the time of disc tearing and are therefore representative of protoplanetary disc conditions. Although the shear viscosity of the retrograde simulation is not shown, the disc is thicker and therefore even better resolved than the simulations discussed in this section.

\subsection{Prograde discs}
\label{sec:prograde}
The thickness of the GW Ori disc is unknown but protoplanetary discs are expected to have $h/r \sim 0.05$. The SPH simulations presented in \citet{kraus2020} set $h/r = 0.02$, which may be unrealistically thin, and would be more prone to disc tearing. However, the simulations of \citet{smallwood2025} did not produce a broken disc for $h/r = 0.05$. Here, we test $h/r = 0.03$ and 0.04, which are reasonable values for protoplanetary discs, at high resolution.

\begin{figure}
    \centering
    \includegraphics[trim=0cm 0cm 8cm 0cm,clip,width=\columnwidth]{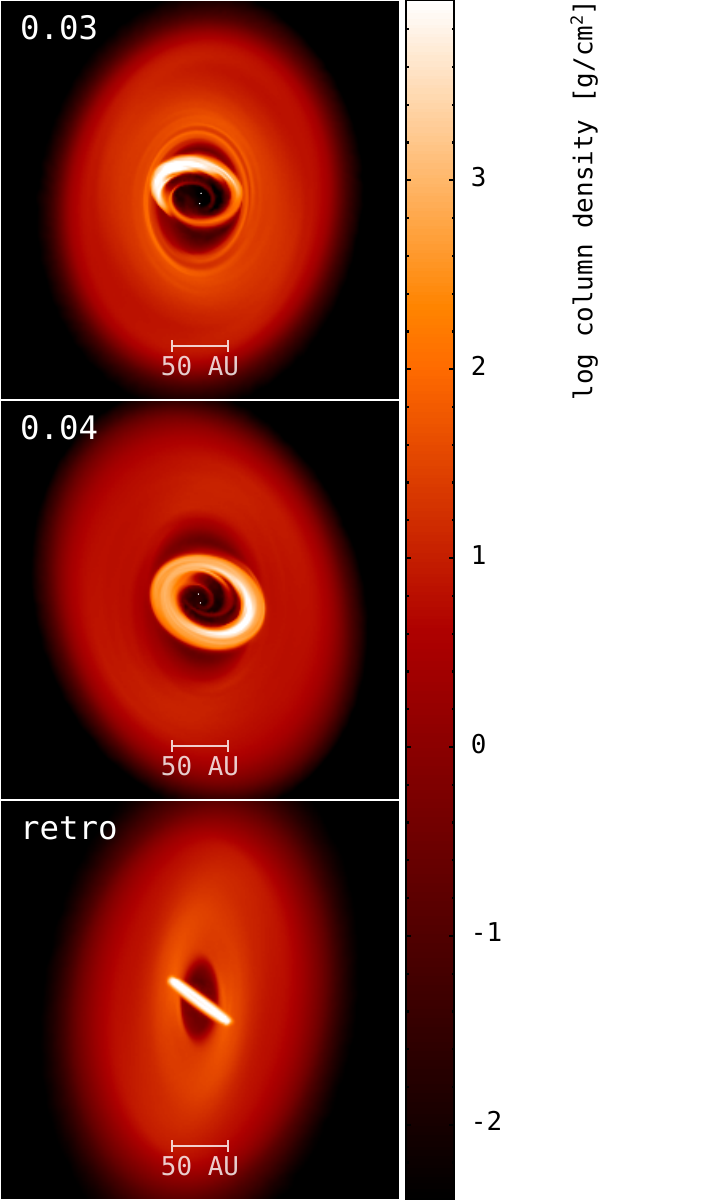}
    \caption{Snapshots from the simulation of the GW Orionis disc with $h/r=0.03$ (top), $h/r=0.04$ (middle) and retrograde $h/r = 0.05$ (bottom), as  viewed on the sky. The snapshots are taken at 260, 420, and 187 binary orbits respectively. All discs are torn by the central binary, forming a misaligned and eccentric ring. The position angle of the outer disc may be rotated to match the observed configuration since we don't know the ``initial'' orientation of the GW Ori disc.}
    \label{fig:densplots}
\end{figure}

\begin{figure*}
    \centering
    \includegraphics[width=0.9\textwidth]{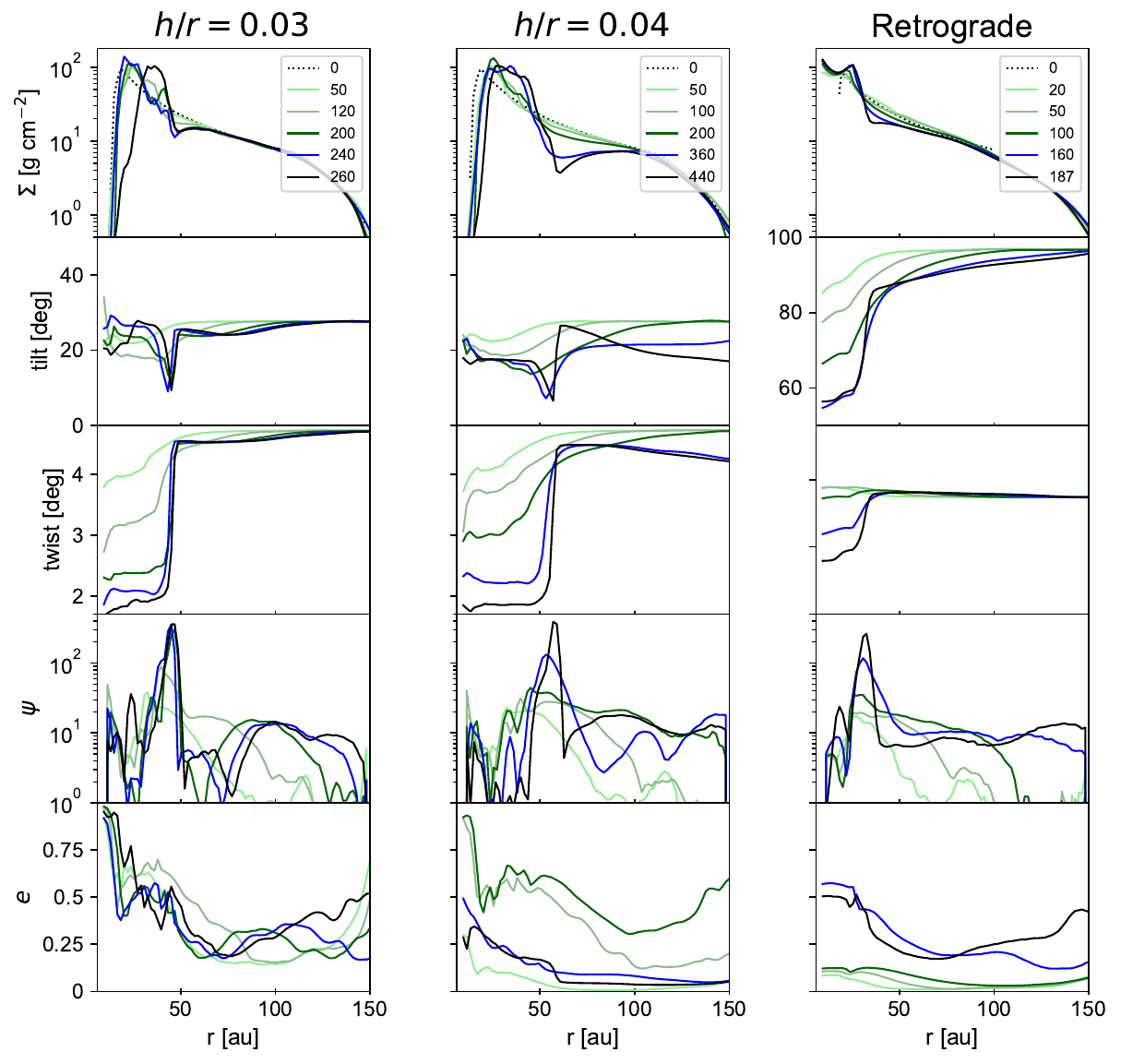}
    \caption{\rev{ Selected snapshots from the simulations of the circumbinary discs. The time of each snapshot is given in binary orbits (AB-C) in the legends and the initial surface density profile is shown by the dotted lines. Panels are surface density ($\Sigma$), tilt, twist, warp amplitude $\psi$, and eccentricity (see section \ref{sec:prograde} for definitions).}}
    \label{fig:warp_profiles}
\end{figure*}

In these new simulations, both discs evolve to become torn by the central binary, as illustrated by the column density plots in Fig.~\ref{fig:densplots}. \rev{The discs are deemed to be broken when there is a sharp radial change in tilt and/or twist that coincides with a local decrease in surface density and spike in the warp amplitude.} The thicker disc takes more than three times as long to tear ($\sim 370$ \rev{binary} orbits vs $\sim 120$). Fig.~\ref{fig:warp_profiles} shows the evolution of the disc and warp properties. The warp amplitude, $\psi$, is described by the change of angular momentum unit vector $\boldsymbol{l}$ with radius $r$ :

\begin{equation}
    \label{eq:warpam}
    \psi = \boldsymbol{r}\left| \frac{\partial \boldsymbol{l}}{\partial r} \right | .
\end{equation}
$\psi$ rises quickly as the breaking instability develops.
The tilt is defined as the angle between the angular momentum vector of the disc and the total angular momentum vector (disc $+$ stars) ${\rm tilt} = \arccos(\hat{l_z}(r))$, where the unit angular momentum vector $\hat{l_z}(r)= l_z(r)/|{\mathbf L}|)$ . Twist is the azimuthal shift defined as ${\rm twist} = \arctan(\hat{l_y}(r)/\hat{l_x}(r))$, where the $x$ and $y$ unit vectors are defined in a similar manner.

From Fig.~\ref{fig:warp_profiles}, we see that the misaligned ring is much wider in the thicker disc ($h/r =0.04$) and the gap between inner and outer discs lies at $r>50$~au, compared to $r\sim 45$~au for the thinner disc. 

The outer regions of the thicker disc have precessed further and developed a different tilt profile to that of the thinner disc, probably because the thicker disc takes longer to break. Before the disc breaks, the outer regions ($r\gtrsim 40$~au) become smoothly warped. After breaking, the tilt decreases in the outer parts of the $h/r=0.04$ disc and the tilt begins to decrease towards the outer edge. The structure of the outer disc differs between the two prograde simulations. There is a surface density contrast of more than an order of magnitude between the inner and outer discs of $h/r = 0.04$, whereas the $h/r=0.03$ disc shows a much smaller difference.

\subsection{Retrograde disc}
\label{sec:retrograde}

The simulation of a retrograde disc is motivated by the degeneracy of rotation direction and inclination in disc observations. Rather than rotating in the same sense as the stellar orbits, the disc could be retrograde and have an inclination different by $180^\circ$, i.e. flipped over, and still have similar observed velocity dust emission maps. The side of the disc that is brighter in infrared scattered light images is typically assumed to be the side facing us due to forward scattering. However, when a disc is strongly warped, the illumination is strongly affected by the position angle of the warp \citep{facchini2018}. The disc is more likely to be prograde with the stellar system but we cannot be certain. Since retrograde discs break more easily \citep{nixon2013}, this scenario is worth modelling. For this simulation, $h/r=0.05$.

The inner edge of the retrograde disc moves closer to the stars than the prograde discs and the disc precesses faster. The inner disc quickly becomes highly misaligned and nearly polar. If the ring reaches a polar configuration it is likely to remain polar \citep{aly2015}. Therefore, this retrograde disc would not present a face-on ring, except for a short time as the ring becomes detached from the outer disc, and so is very unlikely to explain the observations.

\begin{figure}
    \centering
\includegraphics[width=0.9\columnwidth]{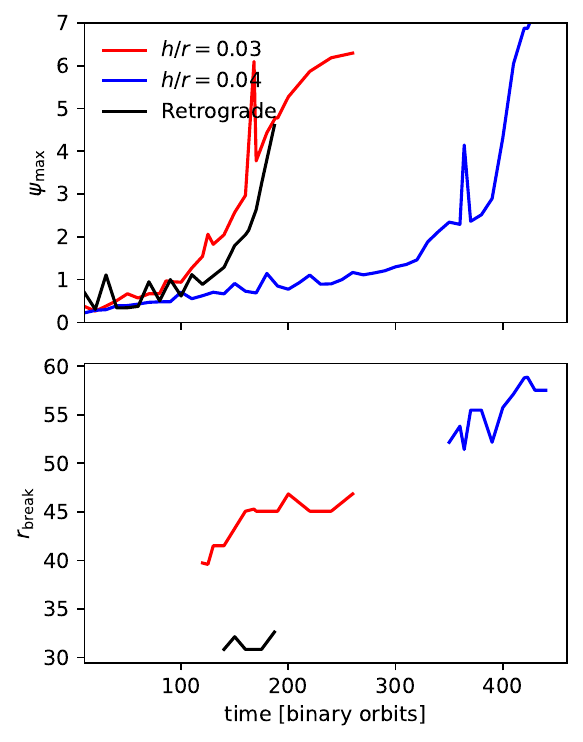}
    \caption{The evolution of the warp amplitude $\psi$ and of the break radius once the disc has broken.}
    \label{fig:psi_evo}
\end{figure}

\begin{figure}
    \centering
    \includegraphics[width=0.9\columnwidth]{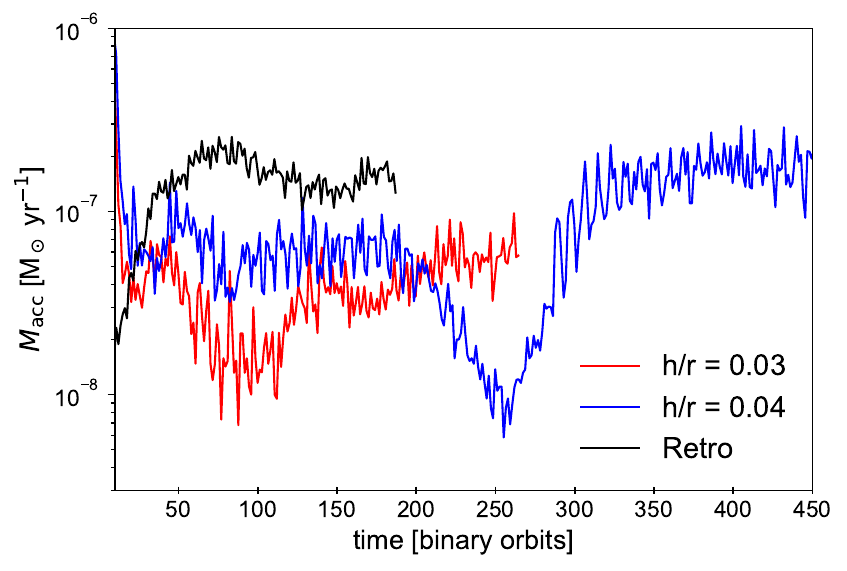}
    \caption{Total accretion rate onto both sink particles averaged over 15.9 years (1.3 binary orbits).}
    \label{fig:macc}
\end{figure}


\section{Analysis and discussion}
\label{sec:discussion}

\subsection{The origin of tearing in GW Orionis}
The simulations here show that stellar torques can cause disc tearing in the GW Ori system. The alternative mechanism proposed by \citet{bi2020} requires a giant planet to clear a gap but, to date, there is no observational evidence of an embedded planet. The planet would need to be several times the mass of Jupiter to clear a gap in the disc \citep{duffell2020} and so should be detectable either directly or through velocity deviations \citep[e.g.][]{pinte2018,teague2018jun}.

Submillimeter continuum and CO observations show hints of a larger-scale `tail' \citep{fang2017,kraus2020}, which \citet{fang2017} attribute to contamination \rev{by a foreground molecular cloud}. Alternatively, this could be evidence of ongoing accretion onto the disc and future observations will be able to address that. Nevertheless, GW Orionis has a large mass and radius which means it likely received a late injection of mass, given the high accretion rate ($3-4 \times 10^{-7}$ M$_\odot$~yr${^{-1}}$, \citealt{calvet2004,fang2014}). Moreover, infrared images \citep{kraus2020} show a highly asymmetric and uneven surface, suggestive of large scale disc motions. Altogether, the characteristics of the misaligned outer disc of GW Orionis point to the outcome of a late accretion event \citep{kuffmeier2021}. \rev{This means that simulations with symmetrical and stable initial conditions cannot tell the whole story but they do provide a useful insight into the system.}


We have established that a retrograde disc does not explain the morphology of the dust rings because the gap is too close to the centre of the disc and because the ring quickly becomes near-polar. The two prograde models give slightly different predictions for the inner ring and gap. The radius of the inner ring of GW Ori is $\sim 45$~au \citep{kraus2020} and this is best matched by the $h/r=0.04$ simulation. The density of the $h/r = 0.03$ inner disc (Fig.~\ref{fig:warp_profiles}) peaks too close to the centre, while the larger breaking radius of $\sim 55$~au for the $h/r = 0.04$ disc allows a thicker inner ring or disc to develop (Fig.~\ref{fig:psi_evo}). Using a grid-based code, \citet{rabago2024} found that a disc with the properties of GW Ori did not break with $h/r=0.075$, which provides a strong upper limit on the disc thickness.
The eccentricity of the ring is observed to be $e=0.3$ \citep{kraus2020} or $e=0.2$ \citep{bi2020}, which is smaller than obtained here for any of the simulated discs. The eccentricity of the inner disc appears to be decreasing as it settles following tearing. The inner disc of the $h/r = 0.04$ model reaches a lower eccentricity but is still $e\gtrsim0.3$.

There is still significant uncertainty in the values of nearly every parameter defining the GW Orionis system so it is possible, of course, that stellar torques alone are insufficient to break a disc with the temperature and density structure of the GW Ori disc. 
\rev{For example, if we consider \textit{pessimistic} orbital parameters instead, we obtain a smaller misalignment between disc and outer binary of $\Phi=23.7^\circ$. Additionally, the choice of $q=0.5$ leads to a flat disc (constant $h/r$) whereas irradiated discs are expected to have a flared structure which would shorten the communication timescale (see \citealt{young2023}).}
However, observations show that the disc is not smooth. As such, the disc tearing mechanism responsible for the inner ring could be assisted by a discontinuity in the disc as proposed by \citet{bi2020,smallwood2021,smallwood2025}. This does not have to be due to a planet. Other possible causes of a sharp change in the radial density and/or pressure include an irregular accretion flow following a late-infall event, or meridional circulation driven by strong poloidal magnetic fields \citep[e.g.][]{hu2022formationofdustrings}.

\subsection{Accretion rate}
The accretion rates for the three models are plotted in Fig.~\ref{fig:macc}. The accretion rate is erratic, varying by factors of 2-3, and seems to settle at $\sim 5 \times 10^{-8}$, $\sim 1.5\times 10^{-7}$ and $\sim 1.5 \times 10^{-7}$ \solmass{}~yr$^{-1}$ for the $h/r = 0.03$, $h/r = 0.04$, and retrograde discs respectively. These values are slightly lower than the observational estimate of $3-4 \times 10^{-7}$ M$_\odot$~yr${^{-1}}$, \citep{calvet2004,fang2014}. The disc mass used in these simulations was estimated by fitting models to the SED, 1.3~mm continuum, and $^{12}$CO data together \citep{fang2017}. The $^{12}$CO line and even the 1.3~mm continuum are optically thick in discs which means that derived masses are underestimated, in some cases by up to an order of magnitude \citep{tung2024}. Increasing the simulated disc mass \rev{(by increasing $\Sigma(r)$)} could well result in higher accretion rates close to the observed value. 

\subsection{Sources of uncertainty and discrepancy with other works}
\label{sec:uncertainty}

In this work, we find that the disc can be torn by the torques exerted by the central stars and this contrasts with some other work. We now examine the reasons for the discrepancies.

\citet{smallwood2025} calculated the initial vertical resolution of their simulations to be ${\langle h \rangle}/{H} = 0.70$. Considering only the quadratic term in equation \ref{eq:alpha_ss} and assuming the authors employed the conventional $\beta_{\rm{AV}}=2$, their equivalent viscosity $\alpha_{\rm{ss}}$ cannot be smaller than 0.04. Moreover, the resolution then worsens as the simulation evolves and the surface density decreases in the warp prior to breaking, further reducing the resolution at the crucial location \citep{drewes2021}. Hence, the simulated discs are more dissipative than protoplanetary discs are expected to be and, since they do not satisfy $ \alpha_{\rm{ss}}\ll h/r$, do not break in the wavelike regime. \rev{The $h/r =0.05$ disc of \citet{smallwood2025} did not break with a misalignment of either $28^\circ$ or $38^\circ$. We were unable to test a disc this thick so we cannot rule out that a $h/r =0.05$ disc is stable with the parameters we used in this work. However, as well as being too thick to break in the viscous regime, the $h/r =0.05$ discs of \citet{smallwood2025} are probably too diffusive for bending waves to propagate through the disc. The latter effect is significant because \citet{young2023} demonstrated that, in wavelike discs, the deformation of the disc at a given radius is due to the combination of the precession torque of the central stars acting locally at that radius and the warp wave driven in the inner disc passing through that radius. Hence, damping the bending waves with excess numerical dissipation reduces the maximum warp amplitude and prevents the disc becoming unstable. }

\rev{Recent works have discussed how the location of the inner edge affects whether the disc breaks \citep{smallwood2021,young2023}. \citet{rabago2024} also pointed out that an initial surface density profile that is artificially too dense at the inner edge would assist breaking due to the resulting increased precession torque. The simulations presented here show the inner edge is dynamic, eccentric, and depends on the orbital phase of the host stars. Accretion streams appear periodically, channelling material inwards. In GW Ori these accretion streams likely feed the circumbinary and circumstellar discs (which are not resolved in our simulations). A dynamic inner edge would explain the observed variability of the SED and of the light curve. 
We find the inner edge to lie $15<r_{\rm{in}}<20$~au, and emphasize that material can flow inside the `edge' to the stars.}

It is difficult to argue how the surface density near the inner edge should be initialised. If the initial inner edge is set further out, given enough time, gas will reach the truncation radius and may pile up, increasing the torque, and result in tearing. This kind of computation is too expensive to be feasible. Additionally, it is not possible to replicate the `real' initial unwarped state of the GW Ori disc because we don't know its history. Analytical estimates of truncation radius assume a settled disc with a steady accretion rate, which is not true of GW Ori.

The accretion rates obtained in the simulations are lower than observed and we suggested that this could be due to an underestimate of the disc mass. An increased disc mass \rev{(and therefore $\Sigma(r)$)} would also increase the precession torque, which would promote disc tearing. Because the disc mass is a key source of uncertainty, better constraints from observations using more recent techniques would be valuable.

\rev{In this work we have approximated the hierarchical triple stellar system as a binary, considering components AB and C with AB located at the barycentre of the A-B orbit. This is the identical approach to other simulations of GW Ori with the exception of \citet{kraus2020}, who modelled the full triple system. As mentioned in \ref{sec:initconds}, \citet{smallwood2021} argue from n-body simulations that the binary approximation is valid but we now compare the angular momentum of the system components to interrogate this further. Based on the orbital parameters of \citet{kraus2020}, we estimate the angular momentum of the inner binary to be $|L_{\rm A-B}|= 1.7\times 10^{53}$~g~cm$^{2}$~s$^{-1}$ and the outer binary to be $|L_{\rm AB-C}| = 6.3\times 10^{53}$~g~cm$^{2}$~s$^{-1}$. The simulated disc here has $|L_{\rm disc}| = 1.4\times 10^{53}$~g~cm$^{2}$~s$^{-1}$. The inner binary contains a significant fraction of the angular momentum of the whole triple system. Fig.~1 of \citet{smallwood2021} compares the secular evolution of a test particle in the disc around the GW Ori triple system to around the approximated binary. The amplitude of the tilt oscillations of the circumtriple test particle is a few degrees larger, which could make all the difference in a marginally stable disc. It is therefore possible that the outcome of a simulation with the full triple system could be slightly different. }

\rev{The above estimates indicate that the angular momentum of the stellar system is similar in magnitude to that of the simulated disc. Consider that we modelled only the inner $150$~au of the $>400$~au disc and also that the disc mass is probably underestimated. In typical disc simulations, $m_{\rm disc} \ll m_{\rm stars}$ but in GW Ori, the disc mass is likely higher. Since analytical estimates of precession and alignment timescales are derived assuming a negligible disc mass, it is probably not meaningful to apply them to the GW Ori system. The evolution of a massive disc is likely different once the evolution of the binary (or triple) due to the disc is considered \citep[see][]{martinlubow2019}. 
The disc and stellar orbits were most likely more misaligned in the past, at which point disc tearing would have been easier. The star cluster formation simulation of \citet{bate2018} generated several protostellar systems in which the disc is tilted within a few 10s~kyr of star formation by ongoing accretion from surrounding gas, consistent with the morphology and timescale of GW Ori ($\sim 1$ Myr). This suggests that what we observe could simply be an expected product of turbulent star formation.}

Observations indicate that the size of the GW Ori gas disc is $R>1000$~au \citep{fang2017} but we simulate only the central $R>150$ au due to computational limitations. This simplification is problematic if the disc breaking occurs after the warp wave reaches the outer edge of the model disc. The warp wave reflects off the edge of the model disc and returns inwards, which could increase the warp amplitude locally beyond the value for an outward propagating wave and/or set up a standing wave. Looking at the evolution of the warp amplitude in Fig.~\ref{fig:warp_profiles}, the warp wave reaches the edge of the $h/r = 0.04$ disc before it breaks. We can compute the communication timescale, the time taken for a bending wave to traverse the disc, from the simulations by estimating the sound speed in $i+1$ annuli, $c_{\rm s,i}$, and integrating over the disc:
\begin{equation}
    t_{\rm{comm}} = \int^{R_{\rm{out}}}_{R_{\rm{in}}} \frac{2}{c_{\rm s}}dr \approx
    \sum_i \left [ (r_{i+1}-r_i) \left ( \frac{1}{c_{\rm s,i}} + \frac{1}{c_{\rm s,i+1}}\right ) \right ].
\end{equation}
The sound speed at radius $r_i$ is ${c_{\rm s,i}} = H_i\Omega_{K,i}$, where the scale height $H_i$ is found from the particle distribution about the midplane.

For the $h/r = 0.04$ disc, $t_{\rm{comm}}\sim 240$ binary orbits, before the disc breaks at $\sim 350$ orbits. The wave reflects inwards, however it will take another $\sim 200$ orbits to reach the location of the break at $\sim 50$~au. For this reason, we expect the result that the $h/r = 0.04$ disc breaks to be valid for the full disc. The $h/r = 0.03$ simulation is unaffected by this issue because the $t_{\rm{comm}}\sim$~360 binary orbits and it breaks after $\sim$~120 binary orbits.

\section{Conclusion}

We have conducted new simulations of the GW Ori disc to investigate whether the stellar torques alone can tear the disc in the wavelike regime. These simulations support the original mechanism proposed by \citet{kraus2020} of disc tearing driven by the stellar torques due to the misalignment between stellar orbits and disc. Comparing the simulations and observations tentatively points to the thicker disc with $h/r =0.04$ being a better match for GW Orionis than $h/r = 0.03$.

This work took optimistic values for the key variables within the ranges of observational uncertainty. A smaller misalignment angle and larger stellar mass ratio could result in a stable warp. Similarly, an initial density profile that is smoothed at the inner edge may prevent sufficient mass building up close enough to the stars for torques to tear the disc. On the other hand, there are differences that could make the disc {\emph{more}} likely to break than in these simulations. These include a non-smooth density profile and a broken temperature profile (due to shadowing) which are probably more representative of the real disc. Mechanisms such as the dust back-reaction on the gas might affect the pressure/density enough locally to assist tearing as well.

The system is complicated and, from what we can tell from observations, the disc hovers near the boundary of a stable warp and tearing instability. Analytical predictions of the behaviour of misaligned discs are necessarily derived from stable initial conditions and smooth accretion rates \rev{, in contrast to the complex morphology of young discs in star-forming environments}. We should accordingly allow some leeway in interpreting them for marginally stable parameters. Even if a smooth disc retains a coherent warp structure, it would only take a small perturbation in the surface density or sound speed profile to break it. We have demonstrated that we cannot robustly rule out tearing by stellar torques alone.

GW Ori remains an exciting system for studying disc evolution and for testing models of misaligned discs. Further observational characterisation of GW Ori is desperately needed to hone hydrodynamical models, in particular to deduce the temperature profile of the disc and to better constrain the gas mass. 

\section*{Acknowledgements}

I am grateful to the anonymous reviewer for their insightful suggestions that helped to improve the paper. I thank Matthew Bate and Chris Nixon for discussions and helpful comments on the manuscript. AKY is grateful for funding from the Royal Society, a Warwick Prize Fellowship, and a UKRI Stephen Hawking Fellowship. This work used the DiRAC Data Intensive service (DIaL2) at the University of Leicester, managed by the University of Leicester Research Computing Service on behalf of the STFC DiRAC HPC Facility (www.dirac.ac.uk). The DiRAC service at Leicester was funded by BEIS, UKRI and STFC capital funding and STFC operations grants. DiRAC is part of the UKRI Digital Research Infrastructure. This work also made use of the Avon HPC cluster operated by the Scientific Computing Research Technology Platform at the University of Warwick. This work made use of {\sc splash} \citep{price2007}, {\sc numpy} \citep{harris2020} and {\sc matplotlib} \citep{hunter2007}.

\section*{Data Availability}

{\sc phantom} is a public code and can be cloned from \url{https://github.com/danieljprice/phantom}. The full set-up files and files for running the simulations will be made available on publication.



\bibliographystyle{mnras}
\bibliography{papers} 


\appendix

\section{\rev{The contribution of eccentricity to the disc evolution}}

\citet{aly2015} demonstrated that the eccentricity of a stellar binary contributes to the precession torque in the circumbinary disc, and can become dominant in some cases. They derive the precession torque due to an eccentric binary on a ring at radius $r$ in the disc with specific angular momentum $\mathbf{l}$ to be:

\begin{equation}
    \mathbf{G_p} = \mathbf{\Theta} \times \mathbf{l},
\end{equation}

with

\begin{equation}
    \mathbf{\Theta} = \frac{3\omega_{\rm K}\mu} {4(1+\mu)^2}
    \frac{a^2}{r^2}
    \left [5e^2 (\mathbf{\hat{l} \cdot \hat{e}}) \mathbf{\hat{e}} - (1-e^2)(\mathbf{\hat{l} \cdot \hat{h}}) \mathbf{\hat{h}} \right ].
\end{equation}
Here $\omega_{\rm K}$ is the orbital angular frequency of the ring and $\mu = M_2/M_1$. The unit vector $\mathbf{\hat{e}}$ points to periastron and $\mathbf{\hat{h}}$ is in the direction of the binary angular momentum vector, i.e. perpendicular to the orbital plane. The term in square brackets gives the relative contributions of the binary eccentricity and binary-disc misalignment to the torque. Note that this depends on the angle between $\mathbf{\hat{e}}$ and $\mathbf{\hat{l}}$ and well as the magnitude of the eccentricity.

For the parameters of GW Ori, we find the components to be $0.049 \mathbf{\hat{e}}-0.750\mathbf{\hat{h}}$. The components of $\mathbf{G_p}$ are then 

\begin{equation}
    \mathbf{\hat{\Theta}} \times \mathbf{\hat{l}} = -0.21 ~\mathbf{\hat{x}} + 0.23 ~\mathbf{\hat{y}} + 0.18 ~\mathbf{\hat{z}},
\end{equation}

\noindent which is offset by $\approx 7^\circ$ from the torque vector when ignoring eccentricity. Considering $e$ increases $\mathbf{|G_p|}$ by $\approx 7$~per cent. The contribution of the eccentricity to the development of the warp in GW Ori is therefore small, but not insignificant.


\bsp	
\label{lastpage}
\end{document}